\documentclass[aps,prl,amsmath,amssymb,preprintnumbers,twocolumn]{revtex4}
\usepackage{amssymb}
\usepackage{color}
\usepackage{graphicx}
\usepackage{subfigure}
\usepackage{epsfig,amssymb,amsmath}
\usepackage[hypertex,linkcolor=red]{hyperref}

\usepackage{color} 

\def\comment#1{}
\def\labell#1{\label{#1}}
\def\section#1{{\par\em #1:--- }}
\def\togli#1{}

\begin{document}

\title{Electromagnetic channel capacity for practical purposes}
\author{Vittorio Giovannetti$^1$, Seth Lloyd$^2$, Lorenzo Maccone$^3$,
  and Jeffrey H. Shapiro$^4$} \affiliation{ \vbox{ $^1$NEST, Scuola
    Normale Superiore and Istituto
    Nanoscienze-CNR, piazza dei Cavalieri 7, I-56126 Pisa, Italy}\\
  \vbox{$^2$Dept.~of Mechanical Engineering, Massachusetts Institute
    of
    Technology, Cambridge, MA 02139, USA}  \\
  \vbox{$^3$Dip.~Fisica ``A.~Volta'', Univ.~of Pavia, via Bassi 6,
    I-27100 Pavia, Italy}\\
  \vbox{$^4$ Research Laboratory of Electronics, Massachusetts
    Institute of Technology, Cambridge, Massachusetts 02139, USA} }
\begin{abstract}
  We give analytic upper bounds to the channel capacity $C$ for
  transmission of classical information in electromagnetic channels
  (bosonic channels with thermal noise). In the practically relevant
  regimes of high noise and low transmissivity, by comparison with
  know lower bounds on $C$, our inequalities determine the value of
  the capacity up to corrections which are irrelevant for all
  practical purposes.  Examples of such channels are radio
  communication, infrared or visible-wavelength free space channels.
  We also provide bounds to active channels that include
  amplification.
\end{abstract}
\maketitle

Shannon \cite{shannon} famously proved that the maximum number of bits
transmitted through a narrowband Gaussian-noise channel is
$C=\log_2(1+S/N)$ for each use of the channel, where $S/N$ is the
signal to noise ratio. At bottom, the noise has a quantum origin, and
the calculation of the capacity requires a quantum description of the
channel.  Accordingly, one of the oldest questions in quantum
information theory is the calculation of channel capacities
\cite{caves,vithol}.

\begin{figure}[hbt]
\begin{center}
(a)\epsfxsize=.44\hsize\leavevmode\epsffile{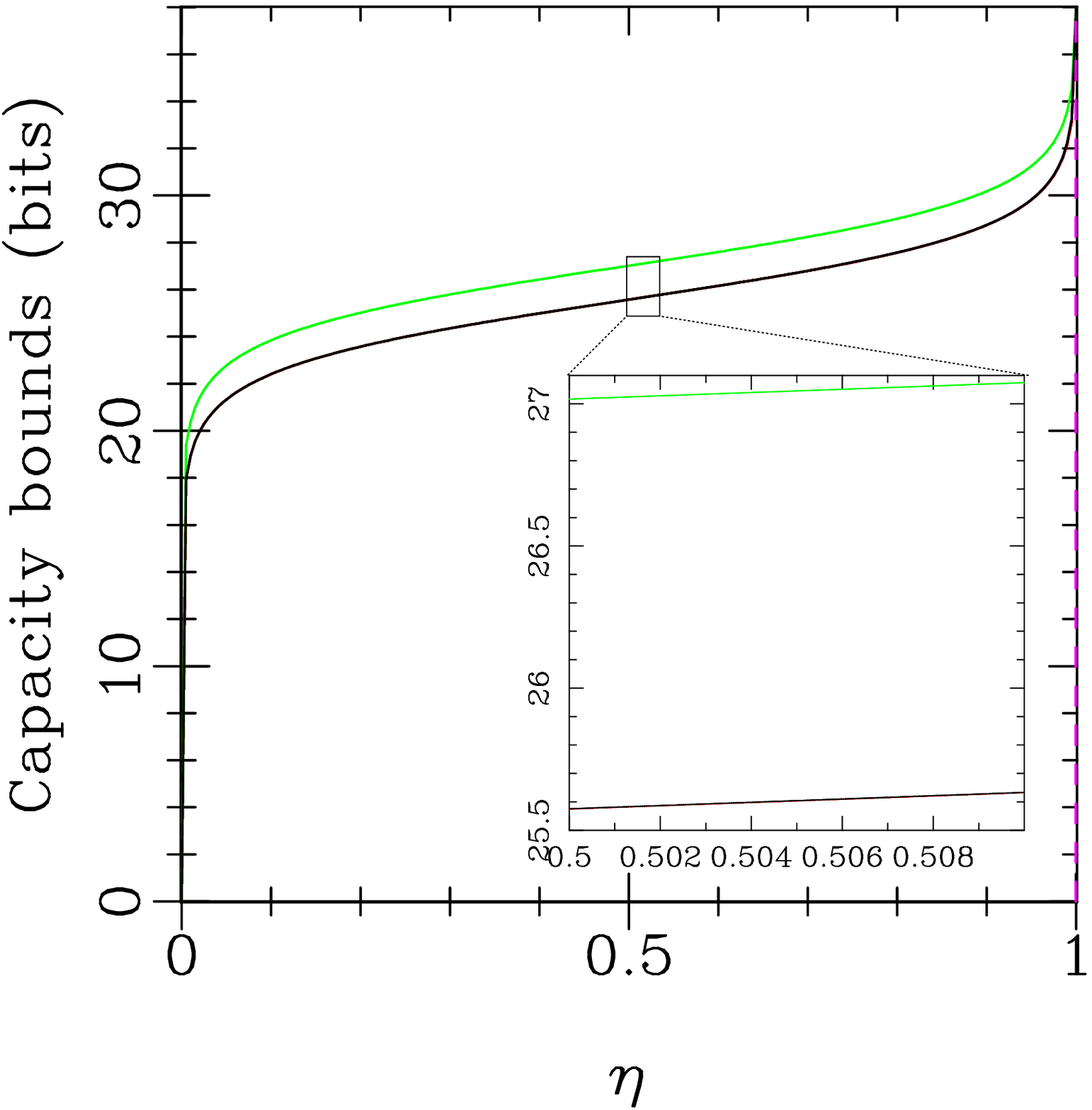}
(b)\epsfxsize=.44\hsize\leavevmode\epsffile{largen-active.eps}
\end{center}
\vspace{-.5cm}
\caption{Plots of the bounds for the passive channel with thermal
  noise (a) and for the amplifier (b) in the practically relevant
  regimes of high thermal noise (large $N$): the upper and lower
  bounds (red and black curves of Figs.~\ref{f:bb}b and \ref{f:bb}c)
  are basically coincident here.\comment{Perhaps here it would be nice
    to put some specific values from some practical channel: I just
    used a 2\,GHz radio link (i.e. a cell phone).}  (a)~Plots of the
  bounds \eqref{b1}, \eqref{b2} and of the bound of Koenig and
  Smith~\cite{konig} as a function of the transmissivity $\eta$ for a
  microwave radio communication, ${\cal E}_\eta^N$, with $N=2000$ at
  room temperature, $\bar N=10^{11}$ for $\sim 1$ mW transmission
  power (assuming that one channel use lasts for one oscillation
  period of the radiation).  In this regime the lower bound \eqref{b1}
  is indistinguishable from our upper bound \eqref{b2} (black curve)
  (see also the magnification in the insert), whereas the gap with
  Koenig and Smith's bound (green curve) is evident. Typical
  transmissivities are very low for these channels (e.g.,~$\eta=0.04$
  for a 14\,dB attenuation).  (b)~Capacity of the amplifier ${\cal
    A}_\kappa^{N}$ as a function of the gain $\kappa$: here the lower
  bound \eqref{b1} is indistinguishable from the upper bound
  \eqref{b5} (black line), whereas the upper bound \eqref{b6} (green
  line) is not useful. Here again $\bar N=10^{11}$ and $N=2000$. }
\labell{f:bg}\end{figure}

Some of the most practically relevant communication channels are
active and passive bosonic channels with thermal noise, e.g.~radio or
infrared-light communication. In this paper we provide upper and lower
bounds for their capacity. In the case of the passive bosonic channel,
our bounds supersede the recent one given by Koenig and Smith
\cite{konig}: in particular in contrast to their bound, for the practically relevant
regimes of large thermal noise or low transmissivities (where each
channel use can convey only small fractions of a bit) our bounds are
sufficiently tight to constitute an expression for the capacity which
is good for practical purposes. These findings are consistent with the
Holevo-Werner conjecture \cite{holevow}, that Gaussian mixtures of
coherent states achieve capacity and that these channels are additive.
In other words, in situations of practical interest, quantum effects
(such as entanglement among subsequent channel uses) do not give any
advantage and a coding alphabet composed of coherent states (e.g.,~the
is output from a maser or laser) achieves capacity. It is important to
stress, however, that there are other regimes in which our
inequalities are not tight: (slight) quantum advantages in low-noise
regimes might be still possible.

We consider two passive channels: the thermal bosonic channel ${\cal
  E}_\eta^N$ that can be modeled by a beam splitter of transmissivity
$\eta$ that mixes the signal with a thermal state with mean photon
number $N$ (the capacity of this channel for $N=0$ is already known
\cite{nostro}), and the classical additive noise channel ${\cal N}_n$
in which the signal is randomly displaced in the complex phase-space
according to a Gaussian probability distribution of variance $n$.  We
also consider phase-insensitive amplifiers ${\cal A}_\kappa^N$ with
gain $\kappa\geqslant 1$, whose additional input mode (required to
ensure the correct commutation relations of the fields) is in a
thermal state of mean photon number $N$. Using a signal of $\bar N$
average photons, with an alphabet of Gaussian-distributed coherent
states, one find the following lower bounds for their capacities
$C$: 
\begin{eqnarray} C({\cal
      E}^N_\eta)&
    \geqslant & g(\eta \bar N+(1-\eta)N)-g((1-\eta)N)\;, \label{b1} \\
    C({\cal N}_n)& \geqslant & g(\bar{N} + n )-g(n)\;,\label{b1b}
    \\ 
    C({\cal A}^N_\kappa) &\geqslant& g(\kappa \bar
    N+(\kappa-1)(N+1))-g((\kappa-1)(N+1)),\nonumber \\
\label{b1c}
\end{eqnarray}
where $g(x):=(x+1)\log_2(x+1)-x\log_2 x$. (A proof of the
Holevo-Werner conjecture would turn these into equalities, but it has
been elusive even after a decade of concerted efforts
\cite{holevow,pra,conj1,jon}.)  The main result of our paper is a
collection of upper bounds which asymptotically match the above lower
bounds in the practically relevant regimes of high noise, or low
transmissivity, or high amplification (see Fig.~\ref{f:bg}). In
particular we show that the following inequalities apply
\begin{eqnarray}
C({\cal E}^N_\eta)&\leqslant& g(\eta \bar N+(1-\eta)N)- g((1-\eta)N-\eta)\;,
\labell{b2}\\
&&\quad\mbox{ for }\eta\leqslant \tfrac{N}{N-1}\;, 
\nonumber\\
C({\cal N}_n)&\leqslant& g(\bar N+n)-g(n-1)\mbox{ for }n\geqslant 1,
\labell{cn1}\\
C({\cal A}^N_\kappa)&\leqslant&g(\kappa \bar N+(\kappa-1)(N+1))-g(N(\kappa-1)-1)\;,
\nonumber\\&&
\quad \mbox{ for }
\kappa\geqslant\tfrac{N+1}N 
\labell{b5}\;.
\end{eqnarray}
 In addition to these simple bounds (which are nonetheless good enough
for many applications) we also derive other, even tighter, bounds in
what follows. All our bounds apply to narrowband channels, but they can be
extended to broadband channels using the variational techniques we
detailed in \cite{nostroshor}. The remainder of the paper is devoted
to the proof of these and of the further bounds.

\section{On Bounds and Conjectures} 
To characterize the channels one can use their action on the state's
characteristic function $\chi(\mu):=\mbox{Tr}[\rho\; e^{\mu
  a^\dag-\mu^* a}]$ ($a$ being the annihilation operator of the mode):
$\chi(\mu)$ is transformed by the channels ${\cal E}_\eta^N$, ${\cal
  N}_n$, and ${\cal A}_\kappa^N$ as (see, e.g.,~Ref.~\cite{pra})
\begin{eqnarray}
  \chi(\mu)&\stackrel{{\cal E}_\eta^N}{\longrightarrow}&\chi(\eta\mu)\;e^{-(1-\eta)(N+1/2)|\mu|^2} \;, 
\labell{chi1} \\
\chi(\mu)&\stackrel{{\cal N}_n}{\longrightarrow}&\chi(\mu)\;e^{- n |\mu|^2}\;,\nonumber\\
\chi(\mu)&\stackrel{{\cal A}_\kappa^N}{\longrightarrow}&\chi(\kappa\mu)\;e^{-(\kappa-1)(N+1/2)|\mu|^2}
\nonumber \;.
\end{eqnarray}

The main difficulty in the calculation of the classical capacity of a
quantum channel $\Phi$ is superadditivity \cite{vithol}: there exist
channels \cite{hastings} in which the alphabet that achieves capacity
must be composed by entangled quantum states that span multiple
channel uses. Accordingly, one must regularize as follows~\cite{vithol}
\begin{eqnarray}
C(\Phi)=\lim_{m\to\infty}C_\chi(\Phi^{\otimes m})/m\:,
\labell{cap}\;
\end{eqnarray}
where $\Phi^{\otimes m}$ indicates $m$ uses of the channel $\Phi$, and
\cite{hsw,sw} \begin{eqnarray} C_\chi(\Psi)=\max_{\{ p_i,\rho_i\}}
 S\Big(\sum_ip_i\Psi[\rho_i]\Big)- \sum_ip_iS(\Psi[\rho_i])
\labell{chi}\;.
\end{eqnarray}
Here $S(\rho):=-\mbox{Tr}\rho\log_2\rho$ is the von Neumann entropy,
$\Psi[\rho_i]$ is the output state from the channel $\Psi$ (that may
represent multiple uses of $\Phi$), and the maximization is performed
over the set of ensembles $\{p_i, \rho_i\}$ formed by density matrices
$\rho_i$ and probabilities $p_i$ that may satisfy some resource
constraint (such as on the average photon number $\bar N$
discussed above). Lower bounds to $C(\Phi)$ can be obtained by
calculating the right hand side of (\ref{chi}) for a specific encoding
alphabet, i.e.~fixing the value of $m$ (say $m=1$) and using a
specific choice of for $p_i$ and $\rho_i$, as was done to obtain the
inequalities \eqref{b1}-\eqref{b1c}. In contrast, an upper bound for
$C(\Phi)$ is provided by
\begin{eqnarray} 
C({\Phi}) \leqslant S_{max}(\Phi) -\lim_{m\rightarrow \infty} { S_{min}(\Phi^{\otimes m})}/{m}\;, \label{upper}
\end{eqnarray} 
where $S_{max}(\Phi)=\max_{\rho} S(\Phi(\rho))$ is the maximum output
entropy for a single channel use [using the same restrictions in the
maximization as in the definition of $C(\Phi)$], and $S_{min}(\Psi) =
{\min_{\rho} S({\Psi}(\rho))}$ is the (unrestricted) minimum output
entropy of the channel $\Psi$. The regularization over $m$ in
\eqref{upper} is required by the superadditivity of the minimum
output entropy~\cite{hastings}, and constitutes the main difficulty in
deriving bounds through (\ref{upper}). However, if $\Phi=\Phi_{EB}$ is
entanglement-breaking~\cite{EBC,holevoeb}, the regularization is
unnecessary~\cite{SHOR}
and (\ref{upper}) can be replaced by
\begin{eqnarray} 
C({\Phi}_{EB}) \leqslant S_{max}(\Phi_{EB}) -
S_{min}(\Phi_{EB})\label{uppereb}\;,
\end{eqnarray} 
 [notice that {\em both}  $S_{min}(\Phi_{EB}^{\otimes m})$
and $C_\chi(\Phi_{EB}^{\otimes m})$ are additive quantities].

For $\Phi={\cal E}_\eta^N$, ${\cal N}_n$, or ${\cal A}_\kappa^{N}$ the
first term on the right of inequality \eqref{upper}, $S_{max}(\Phi)$,
is easily computed by exploiting the fact that the thermal state
maximizes the entropy for fixed average photon number $\bar N$:
\begin{eqnarray}
S_{max}({\cal E}_\eta^N) &=& g(\eta \bar N+(1-\eta)N) \;,  \label{smax} \\ 
S_{max}({\cal N}_n) &=& g(\bar N+n) \;,  \nonumber 
\\
S_{max}({\cal A}_\kappa^N) &=&  g(\kappa \bar N+(\kappa-1)(N+1)) \;.  \nonumber 
\end{eqnarray}  
In contrast, evaluating the second term on the right of inequality
(\ref{upper}) is extremely demanding: the Holevo-Werner conjecture can
be rephrased into a conjecture on the values of $S_{min}(\Phi^{\otimes
  m})$~\cite{pra,AIP,conj1,raul}, which states that the $min$ is
achieved by a vacuum state $|0\rangle^{\otimes m}$.  If this were
true, one could use \eqref{upper} to provide upper bounds that exactly
match the lower bounds \eqref{b1}-\eqref{b1c}. A proof of this is
lacking, but in Ref.~\cite{pra} several bounds were obtained for the
special case of $m=1$: they constrain $S_{min}({\cal E}_\eta^N)$ and
$S_{min}({\cal N}_n)$ close to their conjectured values of
$g((1-\eta)N)$ and $g(n)$, respectively. Using (\ref{uppereb}), such
bounds can be immediately translated into constraints on the capacity
$C$ whenever the maps are entanglement-breaking, i.e.~when $\eta
\leqslant N/(N+1)$ for ${\cal E}_\eta^N$, when $n\geqslant 1$ for
${\cal N}_n$, and when $N\geqslant 1/({\kappa-1})$ for ${\cal
  A}_\kappa^N$~\cite{holevoeb}.  For instance, exploiting this fact,
inequality (\ref{b2}) can be derived by replacing the term
$S_{min}({\cal E}_\eta^N)$ of (\ref{uppereb}) with the single-mode
lower bound A of Ref.~\cite{pra}. More generally, the same approach
exploited in~\cite{pra} can be adapted to the multi-channel use
scenario to construct tight inequalities directly for the quantities
${S_{min}([{\cal E}_\eta^N]^{\otimes m})}/m$ and ${S_{min}([{\cal
    N}_n]^{\otimes m})}/m$.  When substituted into (\ref{upper})
together with the identities~(\ref{smax}) these then translate into a
collection of upper bounds for $C$ that hold beyond the
entanglement-breaking regime detailed above.

\begin{figure}[hbt]
\begin{center}
\epsfxsize=.9
\hsize\leavevmode\epsffile{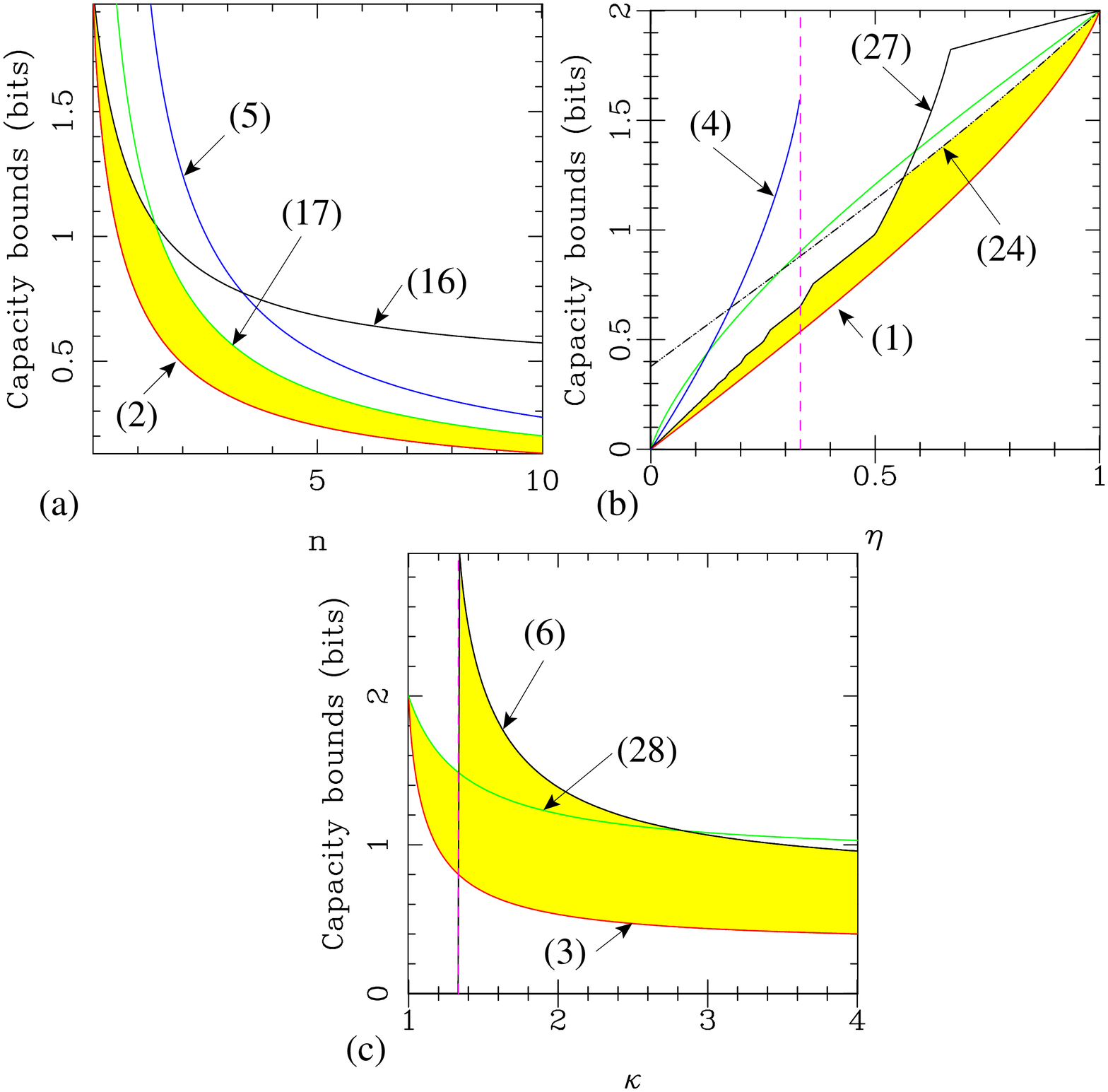}
\end{center}
\vspace{-.5cm}
\caption{Plots of the bounds in regimes that emphasize the gap between
  the lower and the upper bounds (these regimes are typically not
  interesting in practical applications). (a)~Capacity of the Gaussian
  channel ${\cal N}_n$ for $\bar N=1$. Red curve: lower bound
  \eqref{b1c}; blue, black, green curves: upper bounds \eqref{cn1},
  \eqref{cn2}, and \eqref{cn3}, respectively. The yellow area
  emphasizes the gap between the best upper and lower bounds.
  (b)~Capacity of the passive electromagnetic channel ${\cal
    E}_\eta^N$. Red curve: lower bound \eqref{b1}; blue curve: upper
  bound \eqref{b2} (valid only for $\eta\leqslant N/(N+1)$ shown as a
  vertical dashed line where the channel becomes
  entanglement breaking); green curve: Koenig and Smith's bound from
  \cite{konig}; black line: upper bound \eqref{l12}; black dashed
  line: upper bound \eqref{ecn2}. Here $N=0.5$ thermal photons and
  $\bar N=1$ average photons in the signal (which gives
  bits-per-photon for each channel use).  (c)~Plots of the bounds for
  the amplifying channel ${\cal A}_\kappa^N$ with gain $\kappa$.  Red
  curve: lower bound \eqref{b1b}; black curve: upper bound \eqref{b5}
  (valid only for $\kappa\geqslant (N+1)/N$); green curve: upper bound
  \eqref{b6}.  The discontinuity for $\kappa=(N+1)/N$ (vertical dashed
  line) separates the entanglement breaking regime on the right from
  the pure-loss regime on the left.  Here $N=3$ and $\bar N=1$. }
\labell{f:bb}\end{figure}

\section{Bounds for the Additive Classical noise channel ${\cal N}_n$}
As detailed below, the bounds {\em a}, {\em b}, and {\em d} of
Ref.~\cite{pra} for $m=1$ can be generalized to arbitrary $m$ as follows
\begin{eqnarray}
  {S_{min}({\cal N}_n^{\otimes m})}/{m}  &\geqslant &g(n-1) \;,     \quad \mbox{[$\forall n\geqslant 1$]} \label{n1} \\
  {S_{min}({\cal N}_n^{\otimes m})}/{m}  &\geqslant& \log_2(2n+1) \;,  \label{n2} \\
  {S_{min}({\cal N}_n^{\otimes m})}/{m} &\geqslant& 1+ \log_2(n) \;,
  \label{n3}
\end{eqnarray}
whence, using \eqref{upper}, Eq.~\eqref{n1} gives \eqref{cn1}, while
Eqs.~\eqref{n2} and \eqref{n3} respectively give the further bounds
\begin{eqnarray}
C({\cal N}_n) &\leqslant& g(\bar{N} + n) -  \log_2(2n+1) \;,  \label{cn2} \\
C({\cal N}_n) &\leqslant& g(\bar{N} + n) -  1-  \log_2(n) \;, \label{cn3}
\end{eqnarray}
[the generalization of the  bound {\em c} of~\cite{pra} is
not reported here since it  converges to Eq.~(\ref{n2}) for $m \rightarrow
\infty$]. These bounds are compared to the  lower bound (\ref{b1b})
in Fig.~\ref{f:bb}(a): note how the gap  between the upper and lower bounds
closes asymptotically for high noise, $n\rightarrow\infty$. 

The proof of Eq.~(\ref{n1}), and hence of the bound \eqref{cn1}, was
given in Ref.~\cite{AIP} by expanding a generic input state $\rho$ in
terms of its multi-mode Husimi distribution function and applying the
concavity of von Neumann entropy.  An alternative proof follows from
inequality~{\em a} of Ref.~\cite{pra} and from (\ref{uppereb}), using
the fact that the channel ${\cal N}_n$ is entanglement breaking for
$n\geqslant 1$~\cite{holevoeb}.
 
The proof of Eq.~(\ref{n2}) exploits the fact that the von Neumann
entropy is never smaller than the R\'{e}ny{i} entropy of order 2~\cite{ZY,BECK}
i.e.~$S(\rho) \geqslant S_2(\rho):= -\log_2 \mbox{Tr}[\rho^2]$. Thus, for
all input density matrices $\rho$ of $m$ channel uses we have
\begin{eqnarray} 
 S({\cal N}_n^{\otimes m} (\rho)) \geqslant 
 S_2({\cal N}_n^{\otimes m} (\rho)) \geqslant m \, \log_2(2 n+1)\;,
\end{eqnarray} 
where the last inequality follows from the fact that the minimum
R\'{e}nyi entropy of integer order at the output of the channel ${\cal
  N}_n$ is additive and saturated by the vacuum input
state~\cite{pradd}. The bound~(\ref{n2}), and hence \eqref{cn2},
follow by minimizing with respect to $\rho$.
 
The proof of Eq.~(\ref{n3}) closely follows the proof of bound {\em d}
in Ref.~\cite{pra} for $m=1$. Indeed, given a generic pure input state
$|\psi\rangle$, the eigenvalues $\gamma_k$ of the relevant output
state $\rho' ={\cal N}_n^{\otimes m} (|\psi\rangle\langle \psi|)$ can
be expressed as
\begin{eqnarray}
  \gamma_k = 
  \int d^{2 m} \vec{\mu} \; P^{(m)}_n(\vec{\mu}) |\langle \gamma_k|
  D(\vec{\mu}) | \psi\rangle|^2\;,
\end{eqnarray} 
where $|\gamma_k\rangle$ is the corresponding eigenvector of $\rho'$,
$D(\vec{\mu})$ is the $m$-mode displacement operator,
$P_n^{(m)}(\vec{\mu}) := \exp[-|\vec{\mu}|^2/n]/(\pi n)^m$, and the
integral is performed over the $m$-dimensional complex vectors
$\vec{\mu}\in \mathbb{C}^m$. By convexity, for all $z\geqslant 1$ one
can write
\begin{eqnarray}
  \mbox{Tr}[ (\rho')^z] &\leqslant& \sum_k  \int \frac{d^{2 m}
    \vec{\mu}}{\pi^m} \;[ \pi ^m P^{(m)}_n(\vec{\mu})]^z |\langle
  \gamma_k| D(\vec{\mu}) | \psi\rangle|^2\nonumber \\
  &=& {1}/({z n^{z-1}})^m\;,
\end{eqnarray} 
which gives inequality \eqref{n3}, and hence \eqref{cn3}, by
remembering that $S(\rho') = \lim_{z\rightarrow 1^+} { \log_2\mbox{Tr}
  [ (\rho')^z]}/{(1-z)}$~\cite{ZY,BECK}.

\section{Bounds for the Lossy Thermal channel ${\cal E}_\eta^N$} The
bounds (\ref{n1})-(\ref{n3}) can be immediately turned into
inequalities for $S_{min}([{\cal E}_\eta^N]^{\otimes m})$ by
exploiting the compositions rules~\cite{pra} that link ${\cal
  E}_\eta^N$ and ${\cal N}_n$ that also apply to the multi-use
scenario $m>1$. In particular, $[{\cal E}_\eta^N]^{\otimes m} = {\cal
  N}_{(1-\eta)N}^{\otimes m} \circ [{\cal E}_\eta^0]^{\otimes m}$.
Hence, following the same reasoning of Ref.~\cite{pra}, we find
\begin{eqnarray}
  S_{min}([{\cal E}_\eta^N]^{\otimes m})\geqslant S_{min}({\cal
    N}_{(1-\eta)N}^{\otimes m})\;.\label{compo}
\end{eqnarray}
In particular, from (\ref{n2}) and (\ref{n3}) we obtain the multi-use
versions of the bounds B and C of Ref.~\cite{pra}, i.e.,
\begin{eqnarray}
{S_{min}([{\cal E}_\eta^N]^{\otimes m})}/{m}  &
\geqslant& \log_2(2(1-\eta)N+1) \;,  \label{en2} \\
{S_{min}([{\cal E}_\eta^N]^{\otimes m})}{m}  
& \geqslant& 1+  \log_2((1-\eta)N) \;, \label{en3}
\end{eqnarray}
which give rise to the following bounds for the capacity
\begin{eqnarray}
&&C({\cal E}_\eta^N) \leqslant g((1-\eta)\bar{N} + N) - \log_2(2(1-\eta)N+1)  \;,  
\nonumber \\
\label{ecn2}\\
&&C({\cal E}_\eta^N) \leqslant g((1-\eta)\bar{N} + N)  - 1-  \log_2((1-\eta)N)  \;, \nonumber \\
\label{ecn3}
\end{eqnarray}
[the bound obtained from (\ref{n1}) is not reported here as it is
always subsided by the inequality~(\ref{b2})].
Further bounds can be obtained by generalizing the
inequalities E and F of Ref.~\cite{pra}: for all integers $k$, we have
for inequality E:
\begin{eqnarray}
\frac{S_{min}([{\cal E}_\eta^N]^{\otimes m})}{m}   &\geqslant&
\tfrac{k-1}{k} \; g(\tfrac{k}{k-1} (1-\eta) N)
\mbox{ for $\eta\leqslant 1/k$,}\nonumber \\ 
\frac{S_{min}([{\cal E}_\eta^N]^{\otimes m})}{m}
&\geqslant&\tfrac{k-1}{k} \; g\!\left(\tfrac{k}{k-1} \left[(1-\eta) N -\eta
   +\frac{1}{k}\right]\right)\nonumber \\&&\quad
\mbox{for $\eta\geqslant 1/k$,} 
\label{eqnew}
\end{eqnarray} 
and for inequality F:
\begin{eqnarray}
  \frac{S_{min}([{\cal E}_\eta^N]^{\otimes m})}{m}   &\geqslant& 
  \tfrac{k-1}{k} \; g( (1-\eta) N)\nonumber  \\&&
\!\!\!\!\!\!\!\!
+  \tfrac{1}{k} \tfrac{S_{min}([{\cal
      N}_{(1-\eta)N}]^{\otimes m})}{m}    \mbox{ for $\eta\leqslant
    1/k$,} \nonumber\\ 
  \frac{S_{min}([{\cal E}_\eta^N]^{\otimes m})}{m}   &\geqslant& 
  \tfrac{k-1}{k} \; g((1-\eta) N -\eta +\frac{1}{k})\nonumber \\&&
\!\!\!\!\!\!\!\!  + \tfrac{1}{k}
  \frac{S_{min}([{\cal N}_{n'}]^{\otimes m})}{m}  \mbox{ for
    $\eta\geqslant 1/k$}, 
\label{l12}
\end{eqnarray} 
with $n'=(1-\eta)N -\eta + 1/k$ [see the supplementary material for
the proof].  The associated bounds for $C({\cal E}_\eta^N)$ are
obtained by substituting the above expressions into Eq.~(\ref{upper})
together with the identities~(\ref{smax}). In Fig.~\ref{f:bb}(b) we
report the one associated to~(\ref{l12}), together with the
bounds~(\ref{b2}) and (\ref{ecn2}), for a direct comparison with Koenig
and Smith's inequality~\cite{konig}, and with the lower bound
(\ref{b1}). In the low-noise regime of small $N$ a capacity bound was
presented also in \cite{jon}.

\section{Bounds for the Amplifying channel ${\cal A}_\kappa^N$} We now
prove that the capacity of the channel ${\cal A}_\kappa^N$ satisfies
inequality \eqref{b5} and
 \begin{eqnarray}
C({\cal A}^N_\kappa)\leqslant g(\kappa \bar N/[N(\kappa-1)+\kappa]),
 \mbox{ for }N< \tfrac 1{\kappa-1}
\labell{b6}\;,
\end{eqnarray}
see Figs.~\ref{f:bg}(b) and~\ref{f:bb}(c). The proof of inequality
\eqref{b5} can be obtained by adapting the derivation of \eqref{n1}
provided in Ref.~\cite{AIP}. Specifically, the Husimi distribution of
a generic input state $\rho$ for $m$ channel uses is
\begin{eqnarray}
Q({\vec{\alpha}})=\langle\vec{\alpha}|\rho|\vec{\alpha}\rangle/\pi^m
\;, \quad   \rho=\int {d^{2m}\vec{\alpha}}\; Q({\vec{\alpha}}) \;
\sigma(\vec{\alpha})
\labell{husimi}\;,
\end{eqnarray}
where $|\vec{\alpha}\rangle$ is a $m$-mode coherent state and
$\sigma(\vec{\alpha}):= \int \tfrac{d^{2m} \vec{\mu}}{\pi^m}\;
D(\vec{\mu}) \; e^{\vec{\mu}^\dag \cdot \vec{\mu} -\vec{\mu}^T \cdot
  \vec{\mu}^* - \vec{\mu}^\dag \cdot \vec{\mu}/2}$.  The
corresponding output state can then be expressed as $[{\cal
  A}_\kappa^N]^{\otimes m}[\rho]=\int {d^{2m}\vec{\alpha}}\:
Q({\vec{\alpha}}) \; [{\cal A}_\kappa^N]^{\otimes
  m}[\sigma(\vec{\alpha})]$, while the concavity of the output entropy
implies
\begin{eqnarray}
  S( [{\cal A}_\kappa^N]^{\otimes m}[\rho])\geqslant\int
  {d^{2m}\vec{\alpha}}\:Q({\vec{\alpha}})  S( [{\cal
    A}_\kappa^N]^{\otimes m}[\sigma(\vec{\alpha})])
\labell{ka}\;,
\end{eqnarray}
which is meaningful only when $[{\cal A}_\kappa^N]^{\otimes
  m}[\sigma(\vec{\alpha})]$ is a quantum state, i.e.~if
$N(\kappa-1)\geqslant 1$, when $[{\cal A}_\kappa^N]^{\otimes
  m}[\sigma(\vec{\alpha})]$ is an $m$-mode thermal state with average
photon number $N(\kappa-1)-1$ per mode, so that $S([{\cal
  A}_\kappa^N]^{\otimes m}[\sigma(\vec{\alpha})])=m\,g(N(\kappa-1)-1)$.
Substituting it into (\ref{ka}) we find
\begin{eqnarray}
  \tfrac 1m{S_{min}([{\cal A}_\kappa^N]^{\otimes m})}  \geqslant
  g(N(\kappa-1)-1) \;,     \mbox{ for} \; N\geqslant \tfrac
  1{\kappa-1} \label{aa1} 
\end{eqnarray}
which, through~(\ref{upper}), implies~(\ref{b5}).
  
Finally, the proof of the bound \eqref{b6} uses the concatenation ${\cal
  A}_\kappa^N={\cal A}_G^0\circ{\cal E}_\eta^0$ with 
$G=N(\kappa-1)+\kappa$ and $\eta=\kappa/G$, the fact that the capacity is always degraded under channel multiplication, and 
 the fact that $C({\cal E}_\eta^0) = g(\eta
\bar{N})$~\cite{nostro}.

\section{Conclusions} We have given upper and lower
bounds for the classical capacity of important active and passive
bosonic channels, and we have shown that these bounds asymptotically
coincide (yielding the actual capacity) in the regimes of practical
interest, i.e.~for low transmissivity, high thermal noise, or high
amplification.

\section{Acknowledgments} VG acknowledges support from MIUR through FIRB-IDEAS Project  No. RBID08B3FM, LM acknowledges financial support from
COQUIT, and SL and JHS acknowledge support from an ONR Basic Research
Challenge grant.

\newpage

\begin{widetext}
\subsection*{Supplemental Material}

Here we provide explicit derivations of the inequalities~(\ref{eqnew}) and (\ref{l12}).  

\section{Proof of Eq.~(\ref{eqnew})} This inequality can be easily
obtained by generalizing to the multimode scenario the beam-splitter
decomposition of the channel ${\cal E}_\eta^N$ detailed in the
Appendix D1 of Ref.~\cite{pra} [the same decomposition was also
exploited in Ref.~\cite{conj1}].  Consider first the case $\eta =1/k$
with $k$ an integer [generalization to arbitrary $\eta$ will given
later].  The basic idea is to express the transformation induced by
the map ${\cal E}_{1/k}^N$ in terms of a sequence of $k-1$
beam-splitter interactions that couple the incoming signal mode state
$\rho$ with $k$ independent bosonic thermal baths states $\rho_{th}$
characterized by the same photon number $N$. As discussed in
Ref.~\cite{pra} this can be done in such a way that local observers
located at each of the $k$ outputs of the array will receive [up to an
irrelevant local unitary transformation] the {\em same} output signal
${\cal E}_{1/3}^N[\rho]$. For instance for $k=3$ this can be obtained
by setting the transmissivity of the first beam splitter equal to
$\eta_1=2/3$ and the second one to $\eta_2=1/2$. The same construction
clearly can be applied to channel $[{\cal E}_{1/k}^N]^{\otimes m}$ of
the $m$-channel use scenario by repeating the decomposition for each
channel independently. An example of the resulting scheme for $k=3$
and $m=2$ is shown in Fig.~\ref{f:figure1}: here $A_1$ and $A_2$
represent the two channel inputs that in principle can be loaded with
a non separable state $\rho$; $B_1$, $C_1$ are instead the two thermal
bath modes needed to represent the first channel use, while $B_2$ and
$C_2$ are those associated with the second channel use [all of them
being initialized in thermal states having average photon-number $N$].
In this extended configuration one can easily verify that the two-mode
states at the ports $A_1'A_2'$, $B_1'B_2'$, and $C_1'C_2'$ of the
figure are all unitarily equivalent to the density matrix $[{\cal
  E}_{1/3}^N]^{\otimes 2}(\rho)$ [in other words, up to local unitary
transformations, each one of those output couples yields a unitary
dilation~\cite{vithol} of the same channel $[{\cal
  E}_{1/3}^N]^{\otimes 2}$]. This in particular implies that the
associated output entropies must be identical, i.e.  $S(A_1'A_2') =
S(B_1'B_2') =S(C_1'C_2')=S([{\cal E}_{1/3}^N]^{\otimes 2}(\rho))$.
Exploiting the sub-additivity of the von Neumann entropy~\cite{vithol}
we can hence write
\begin{eqnarray}
  S(A_1'A_2'B_1'B_2'C_1'C_2')\leqslant 3 S([{\cal E}_{1/3}^N]^{\otimes 2}(\rho))\;,\label{sss}
\end{eqnarray} 
where $S(A_1'A_2'B_1'B_2'C_1'C_2')$ is the entropy of the joint state
at the output of the device.  Observing that the transformation [i.e.,
the beam-splitter couplings] that takes the input modes of the system
$A_1A_2B_1B_2C_1C_2$ to their associated output
$A_1'A_2'B_1'B_2'C_1'C_2'$ configuration is unitary, we can then
identify $S(A_1'A_2'B_1'B_2'C_1'C_2')$ with the input entropy
$S(A_1A_2B_1B_2C_1C_2)$.  The latter can easily be computed by
noticing that the incoming state is just a tensor product of $\rho$
with $m(k-1)=4$ bosonic thermal states with mean photon-number $N$,
i.e., $S(A_1A_2B_1B_2C_1C_2) = S(\rho) + 4 g(N)$. Substituting this
into Eq.~(\ref{sss}) we finally get
\begin{eqnarray}
  S([{\cal E}_{1/3}^N]^{\otimes 2}(\rho))\geqslant S(\rho) + \frac{4}{3} g(N)\geqslant \frac{4}{3}g(N)\;.
\end{eqnarray} 
The same argument can be easily repeated for arbitrary $m$ and $k$
integers: in this case, we use $m(k-1)$ local bath modes organized in
$m$ parallel rows, each containing $k-1$ beam-splitter transformations
whose transmissivities are determined as in Ref.~\cite{pra}. Similarly
to the case explicitly discussed above, an inequality for $S([{\cal
  E}_{1/k}^N]^{\otimes m}(\rho))$ can be obtained via sub-additivity
by grouping the $m k$ output modes into $k$ subsets of $m$ elements
each.  The resulting expression is
\begin{eqnarray}\label{ffg11}
S([{\cal E}_{1/k}^N]^{\otimes m}(\rho))\geqslant  m \; \frac{(k-1)}{k}\;  g(N)\;.
\end{eqnarray} 
Generalization of this inequality to $\eta\leqslant 1/k$ can finally
be obtained along the same lines used in Ref.~\cite{pra} by exploiting
the following composition rules
\begin{eqnarray} [{\cal E}_{\eta_2}^{N_2}]^{\otimes m} \circ [{\cal
    E}_{\eta_1}^{N_1}]^{\otimes m} = [{\cal
    E}_{\eta_1\eta_2}^{N'}]^{\otimes m}\;,
\end{eqnarray}
which is a trivial multi-mode generalization of the identity (19) from
\cite{pra} [here $N' = [\eta_2(1-\eta_1) N_1 + (1-\eta_2) N_2]
/(1-\eta_1\eta_2)$].  The reader can check that the resulting
expression coincides with the first part of the
inequality~(\ref{eqnew}).  Similarly, Eq.~(\ref{ffg11}) can be used to
induce a bound for $\eta \geqslant 1/k$ by following the same line of
reasoning presented in~\cite{pra} while exploiting the composition
rule
\begin{eqnarray} [{\cal E}_\eta^N]^{\otimes m}= [{\cal
    E}_{\eta'}^{N'}]^{\otimes m}\circ [{\cal
    A}_{\eta/\eta'}^0]^{\otimes m}\;, \qquad \mbox{for $\eta \geqslant
    \eta'$,}
\end{eqnarray} 
which is the $m$-mode counterpart of the identity~(B3) of~\cite{pra}. The resulting inequality yields the second part of~(\ref{eqnew}). 

\begin{figure}[t]
\begin{center}
\epsfxsize=.3\hsize\leavevmode\epsffile{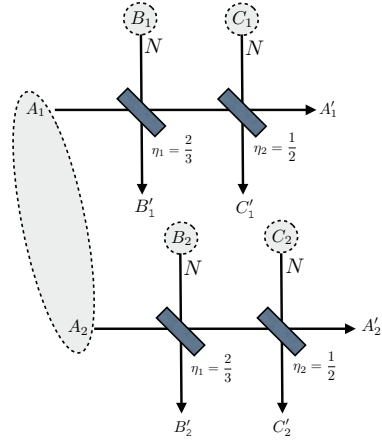}
\end{center}
\vspace{-.5cm}
\caption{Beam splitter-decomposition scheme for the channel $[{\cal
    E}_\eta^N]^{\otimes m}$ with $\eta =1/3$ and $m=2$. Thermal states
  of mean photon-number $N$ are injected at the input ports $B_1$,
  $C_1$, $B_2$, and $C_2$. }
\labell{f:figure1}\end{figure}

\section{Proof of Eq.~(\ref{l12})}
The $m=1$ version of this inequality was derived in \cite{pra}, by
exploiting a beam-splitter array obtained by applying a channel ${\cal
  N}_n$ at each of the output ports of the scheme used to derive the
$m=1$ equivalent of Eq.~(\ref{eqnew}) [see Fig. 12 of ~\cite{pra}].
As for Eq.~(\ref{eqnew}), the main difficulty in applying the same
argument to arbitrary $m$ is generalizing such an array to the
multi-mode case scenario and properly grouping the corresponding
output ports. This can be done as sketched in Fig.~\ref{f:figure2}:
i.e., adding ${\cal N}_n$ to each of the ports in Fig.~\ref{f:figure1}
and by keeping the same grouping scheme as before.  With this guidance
the reader can now closely follow the same derivation given in
Ref.~\cite{pra} [the steps are rather cumbersome, but basically
coincide with those we have discussed in the previous section].

\begin{figure}[t]
\begin{center}
\epsfxsize=.3\hsize\leavevmode\epsffile{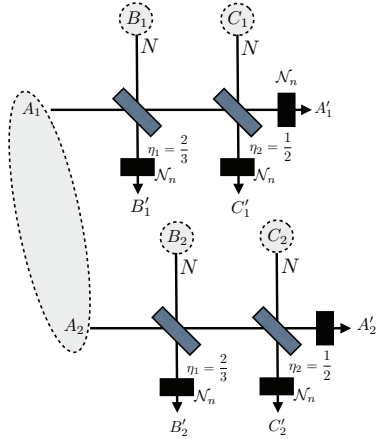}
\end{center}
\vspace{-.5cm}
\caption{Beam-splitter array used to derive Eq.~(\ref{l12}), depicted
  for $m=2$. }
\labell{f:figure2}\end{figure}

\end{widetext}

\end{document}